\documentclass[twocolumn,english,prl,superscriptaddress]{revtex4-2}

\usepackage{graphicx}
\usepackage{dcolumn}
\usepackage{bm}

\usepackage[utf8]{inputenc}
\usepackage[T1]{fontenc}
\usepackage{booktabs, array, mathptmx, float, tabularx, booktabs, lipsum, amsmath,multirow}
\usepackage{siunitx, xcolor}
\usepackage[version=4]{mhchem}
\graphicspath{{figs/}{figsgaoerb/}} 
\usepackage[colorlinks,linkcolor=blue,anchorcolor=blue,citecolor=blue]{hyperref}
\usepackage{amssymb}
\usepackage{ragged2e}

\usepackage{amsmath,amssymb,amsfonts}
\usepackage{CJK}
\usepackage{tikz}
\usepackage{pgfplots}
\usetikzlibrary{matrix,fit}
\usepackage{subfig}
\usepackage{caption}
\usepackage{mathrsfs}
\usepackage{float}
\usepackage{pgflibraryarrows}
\usepackage{lipsum} 
\usepackage{pgflibrarysnakes}

\begin{document}

\renewcommand{\raggedright}{\leftskip=0pt \rightskip=0pt plus 0cm}
\captionsetup[figure]{name={FIG.},labelsep=period}

\title{Successive Phase Transition in Higher-order Topological Anderson Insulators}



\author{Aodong Li}\thanks{These authors contributed equally to this work.}
\affiliation{School of Science and Engineering, The Chinese University of Hong Kong, Shenzhen, 518172, China}
\author{Bingcong Xu}\thanks{These authors contributed equally to this work.}
\affiliation{Wuhan National Laboratory for Optoelectronics and School of Optical and Electronic Information, Huazhong University of Science and Technology, Wuhan 430074, China}
\author{Biye Xie}
\email[]{xiebiye@cuhk.edu.cn}
\affiliation{School of Science and Engineering, The Chinese University of Hong Kong, Shenzhen, 518172, China}



\begin{abstract}
Disorder, traditionally believed to hinder the propagation of waves. has recently been shown to prompt the occurrence of topological phase transitions. For example, when disorder strength continuously increases and surpasses certain critical value, a phase transition from topologically trivial to nontrivial insulating phases occurs. However, in the parameter domain of the nontrivial phase, whether there exists a finer phase diagram that can be further classified by different disorder strengths is still unclear. Here we present a successive topological phase transition driven by the disorder strength in a higher-order topological insulator with long-range couplings. As the strength of the disorder gradually increases, the real-space topological invariant of the system undergoes a consecutive change from 0 to 4, accompanied by the stepped increase in the number of boundary-localized corner states. Our work opens an avenue for utilizing disorder to induce phase transitions among different higher-order topological insulators.

\end{abstract}


\maketitle

{\it{Introduction}}.--- Disorder, widely existed in solid-state materials, plays a pivotal role in the occurrence of many quantum phenomena such as the branch flow of waves~\cite{FW1,FW2,FW3} and the Anderson localization (AL)~\cite{AL1,AL2,AL3}. Concretely, when AL occurs, the diffusive motion of electrons in solids is broken down by disorders, leading to the localization of electronic wavefunctions~\cite{AL4}. The AL is not unique in electronic materials but has also been observed in many classical waves such as photonics~\cite{AL5,AL6} and acoustics~\cite{AL7,AL8}. Interestingly, waves may behave differently to disorder strengths in different dimensions~\cite{AL9}. For example, in one dimension, an arbitrary weak disorder can trigger the occurrence of AL~\cite{AL4} while AL can only happen with a strong disorder in higher dimensions~\cite{AL9}. The study of the interplay between disorder and waves has significantly deepened our understanding of condensed matter physics and stimulated the development of many other physical branches such as high-temperature superconductors~\cite{AL10} and advanced lasering technology~\cite{AL11}. 

In the meantime, topological phases of matter have been extensively studied in the past few decades, well-known for their unique robustness against disorders~\cite{TI1,TI2,TI3,TI4}. Even with topological bandgap protections, under sufficiently strong disorders, the relevant topological phases can be broken and all states are localized~\cite{TI5}. However, counterintuitively, a reverse transition may occur as an increasing weak disorder can drive a phase transition from a topologically trivial phase to a non-trivial phase, denoted as topological Anderson insulators (TAIs)~\cite{TAI1,TAI2}. This phenomenon has been widely studied in one-dimensional (1D) systems~\cite{TAI3,TAI4,TAI5,TAI6,TAI7,TAI8} and has lately been proved in many wave platforms, such as photonics~\cite{TAI9,TAI10,TAI11}, acoustics~\cite{TAI3} and cold atomic wires~\cite{TAI4}. Recently, TAIs have been found in two dimensions, for example, Shunqing Shen et al~\cite{QTAI1}, prove that in two dimensions, disorder can drive a higher-order topological phase (HOTP) transition from a trivial phase to a quantized quadrupole  Anderson insulator~\cite{QTAI2,QTAI3}. As ensured by the higher-order bulk-boundary correspondence, under the open boundary condition, waves are found to be localized at 0D corners. HOTPs have recently been discovered as a family of topological phases that possess lower-dimensional boundary states~\cite{HOTI1,HOTI2,HOTI3}. However, it is still unclear whether such nontrivial HOTPs can be further classified by different disorder strengths and whether certain disorders can induce more corner states at one corner structure. Answering these questions will not only benefit to enrich the family of disordered HOTPs but also promote the potential application of lower-dimensional disordered topological devices.

In this work, we present a successive topological phase transition as the disorder increases continuously in a 2D HOTP. By introducing long-range hopping terms and four different intracell hopping terms between lattice sites in the 2D Benalcazar-Bernevig-Hughes (BBH) model~\cite{HOTI1}, we construct the chiral symmetric higher-order topological Anderson insulators (HOTAIs) with sublattices that respond differently to the disorder strength. When disorders are introduced homogeneously in certain hopping terms of the lattice, the critical values for the HOTAI phase transition are distinct for different sublattices. Therefore, when such homogeneous disorder strength in the lattice continuously increases, the topological phase transition of each sublattice consecutively occurs. We prove this successive topological phase transition by directly calculating the real-space topological invariants and the demonstration of higher-order bulk-boundary correspondence.  Our results uncover the diverse topologically nontrivial phases induced by disorders and provide a flexible way to control the number of topologically protected corner states in one corner structure.

{\it{Results}}.--- We start by considering a tight-binding model with chiral symmetry as shown in Fig. 1(a). The lattices are parted into two sets of sublattices (A, C and B, D) [see Fig. 1(b)]. The Hamiltonian of this model is as following~\cite{RTI1}:
\begin{equation}
H=\begin{bmatrix}
  0&h \\
  h^\dagger &0
\end{bmatrix},
\label{eq:one}
\end{equation}
Here chiral symmetry is manifested as  $\Pi H{\Pi ^{ - 1}} =  - H$, where $\Pi  = \left[ {\begin{array}{*{20}{c}}
I&0\\
0&- I
\end{array}} \right]$ is the chiral operator and $I$  is the identity matrix. In the momentum space $h$ can be expressed as
\begin{equation}
h(k_x,k_y)=\begin{bmatrix}
  \beta _{1} & \kappa _{12} & \kappa _{13} & 0\\
  \kappa _{12} & \beta _{2} & 0 & \kappa _{24}\\
  \kappa _{13} & 0 & \beta _{3} & \kappa _{34}\\
  0 & \kappa _{24} & \kappa _{34} & \beta _{4}
\end{bmatrix}
\label{eq:two}
\end{equation}
with ${\beta _j} = \left[ {\begin{array}{*{20}{c}}
{ - {v_{xj}} - {u_{xj}}{e^{ - 2i{k_x}}}}&{{v_y} + {u_y}{e^{ - 2i{k_y}}}}\\
{{v_y} + {u_y}{e^{2i{k_y}}}}&{{v_{xj}} + {u_{xj}}{e^{2i{k_x}}}}
\end{array}} \right]$ (${j = 1,2,3,4}$), ${\kappa _{12}} = {\kappa _{34}} = \left[ {\begin{array}{*{20}{c}}
{{w_x}{e^{ - i{k_x}}}}&0\\
0&{{w_x}{e^{i{k_x}}}}
\end{array}} \right]$ and ${\kappa _{13}} = {\kappa _{24}} = \left[ {\begin{array}{*{20}{c}}
0&{{w_y}{e^{ - i{k_y}}}}\\
{{w_y}{e^{ - i{k_y}}}}&0
\end{array}} \right]$. Here ${v_{xj}}$ (${v_y}$) is the intracell hopping, ${w_x}$ (${w_y}$) is the intercell hopping, ${u_x}$ (${u_y}$) is the long-range hopping, and ${k_x}$ (${k_y}$) is the wave vector along $x$ ($y$) direction. Under the structure of $H$ described above, the eigenstates can be expressed as combinations of two sublattices: $\left| {\psi_n} \right\rangle=(1 /\sqrt 2){\left[ {\psi _n^{AC},\psi _n^{BD}} \right]^T}$, where $\psi _n^{AC} = {[\psi _n^{{A_1}},\psi _n^{{C_1}},\psi _n^{{A_2}},\psi _n^{{C_2}}, \ldots ]^T}$ is normalized states that exist only in the A and C subspaces, and $\psi _n^{BD} = {[\psi _n^{{B_1}},\psi _n^{{D_1}},\psi _n^{{B_2}},\psi _n^{{D_2}}, \ldots ]^T}$ is normalized states that exist only in the B and D subspaces. If the long-range hoppings are zero and all the intracell (intercell) hoppings are the same as each other. Then this tight-binding model is the well-known BBH model in which a quantized quadrupole insulating phase can emerge when the intracell hopping is smaller than the intercell hopping~\cite{HOTI1}. When disorder is introduced in the BBH model, a phase transition from trivial to higher-order topological phases is presented with a deformation of the phase diagram from the clean limit which is nontrivial in the sense that the topological phase regime can expand due to disorder in certain parameters space~\cite{TAI2}. We here emphasize that the disorder-induced HOTPs in BBH model is in ${\mathbb{Z}_2}$ class.

\begin{figure}
\includegraphics{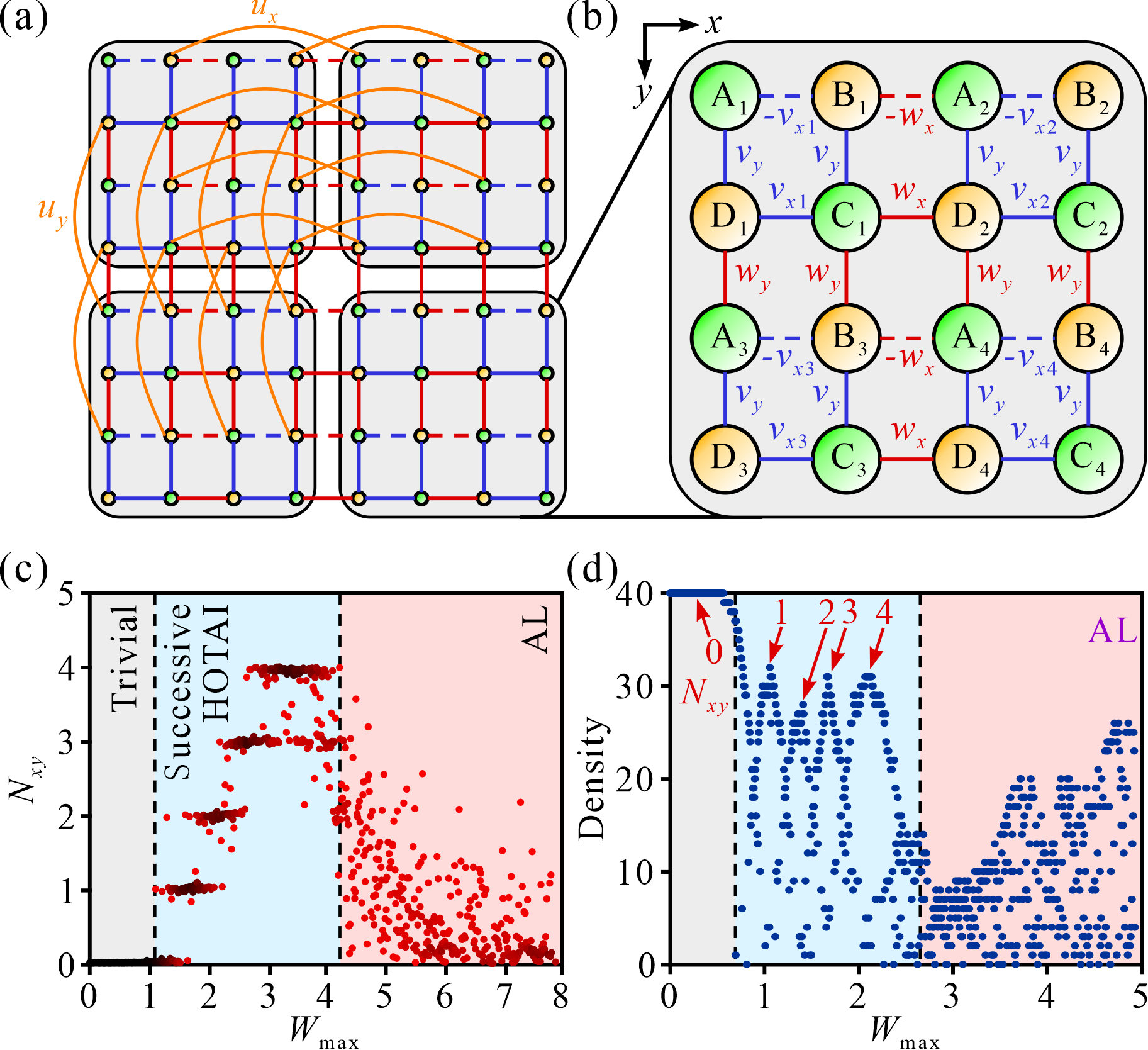}%
\caption{\justifying Schematic depicting the tight-binding model used. The blue, red and orange line denote the intracell, intercell and long-range hopping terms, respectively. The dashed lines represent hopping terms with negative signs such that each plaquette has a uniform flux of $\pi$. Not all long-range hoppings are shown for clarity. (b) Enlarged local image of Fig. 1(a) to show the difference of the hopping terms. The sublattices with opposite chiral charge are distinguished by green and yellow underpaintings. (c) The topological invariant ${N_{xy}}$ versus disorder. The depth of color represents the density of points. The light gray, blue and red regions denote the trivial, successive HOTAI and AL phases, respectively. (d) The density of points versus disorder. The arrow represents the value of ${N_{xy}}$ at each peak.\label{}} 
\end{figure}

To investigate whether the disorder can induce more diverse nontrivial HOTPs, we here introduce non-zero long-range hopping terms. Moreover, four different ${v_{xj}}$, ${j=1,2,3,4}$ are chosen to further construct four sublattices in the Hamiltonian(see Fig. 1(b)). When disorder is introduced homogeneously on the couplings ${v_{xj}}$, these four sublattices respond differently to the same disorder strengths and the successive topological phase transition occurs. To accurately characterize such phase transition, We need to construct a real-space topological invariant as~\cite{LTI1}
\begin{equation}
N_{x y}=\frac{1}{2 \pi i} \operatorname{Tr} \log \left(\bar{Q}_{x y}^{A C} \bar{Q}_{x y}^{B D \dagger}\right)  
\label{eq:three}
\end{equation}
where $\bar Q_{xy}^S = U_S^\dag Q_{xy}^S{U_S}$ for $S = AC$ or $BD$.  ${U_S}$ can be obtained through the singular value decomposition of $h$ by $h = {U_{AC}}\Sigma U_{BD}^\dag $ with ${U_S} = \left[ {\psi _1^S,\psi _2^S, \ldots ,\psi _N^S} \right]$ ($S = AC$ or $BD$) and $\Sigma $ is a diagonal matrix containing the singular values. $Q_{xy}^S$ is the sublattice multipole moment operators defined as $Q_{xy}^{S}=\sum_{\mathbf{R} ,\alpha \in S} \left | \mathbf{R} ,\alpha \right \rangle \mathrm{exp} \left ( -i\frac{2\pi xy}{L_{x}L_{y}} \right )\left \langle \mathbf{R} ,\alpha \right | $, where $L_{x}$ ($L_{y}$) is the number of unit cells along $x$ ($y$)  direction, ${\mathbf{R}} = (x,y)$ represents the position coordinates of each unit cell, $\left| {{\mathbf{R}},\alpha } \right\rangle  = c_{{\mathbf{R}},\alpha }^\dag \left| 0 \right\rangle $, and $c_{{\mathbf{R}},\alpha }^\dag $ creates an electron at sublattice $\alpha $  of unit cell  ${\mathbf{R}}$.

With a well-defined real-space topological invariant  ${N_{xy}}$, we next introduce disorder into the model presented in Fig. 1(a) and investigate the phase transition of the system. The disorder is present on the intracell hopping terms along $x$ direction: ${v_{xj}} = {\bar v_{xj}} + W$ with the random function $W$  distributed uniformly within the interval $[ - W_\mathrm{max},W_\mathrm{max}]$ without correlation. Here we choose ${\bar v_{x1}} = 1.01$, ${\bar v_{x2}} = 1.09$, ${\bar v_{x3}} = 1.13$, ${\bar v_{x4}} = 1.23$, ${v_y} = 0.75$, ${w_x} = {w_y} = 0.1$, ${u_x} = {u_y} = 1$, ${L_x} = 150$ and ${L_y} = 25$. The calculated ${N_{xy}}$ under increased disorder strength $W_\mathrm{max}$  is shown in Fig. 1(c). When the disorder is weak,  $N_{xy}=0$ and the system is in topologically trivial phase. As the disorder strength increases, ${N_{xy}}$ stepwise increases from 0 to 4, representing successive phase transitions. When $W_\mathrm{max}$ further increases, ${N_{xy}}$ is no longer quantized as excessive disorder disrupts the effective hopping between unit cells and the system transit to AL phase. In the HOTAI phase, few points are scattered with ${N_{xy}}$ being not integers, which is caused by the insufficient size used in the computational model and resulting in numerical calculation errors. As limited by the finite-size lattice and the parameters, the phase transition does not occur in a cliff-like manner. For example, during the process as ${N_{xy}}$ changing from 0 to 1, it will repeatedly take the two values. As $W_\mathrm{max}$ increases, ${N_{xy}}$ will have a higher probability of taking the value of 1, while the probability of taking the value of 0 will decrease. To demonstrate the phase transition more clearly, we calculated the density of points with the same ${N_{xy}}$ within the range from $-$0.2 to 0.2 near a certain $W_\mathrm{max}$ (due to calculation errors, ${N_{xy}}$ are considered to be the same if the difference between them is small than 0.2). The results presented in Fig. 1(d) show four obvious peaks in the topological nontrivial phase at $W_\mathrm{max} = $ 1.72, 2.29, 2.72 and 3.41, with ${N_{xy}=}$ 1, 2, 3 and 4, respectively (the interval between the values of $W_\mathrm{max}$  is 0.01, so the maximum density is 40). The peak that occurs when $W_\mathrm{max}$ continues to increase is caused by AL and ${N_{xy}}$ tends to be 0. 

\begin{figure}
\includegraphics{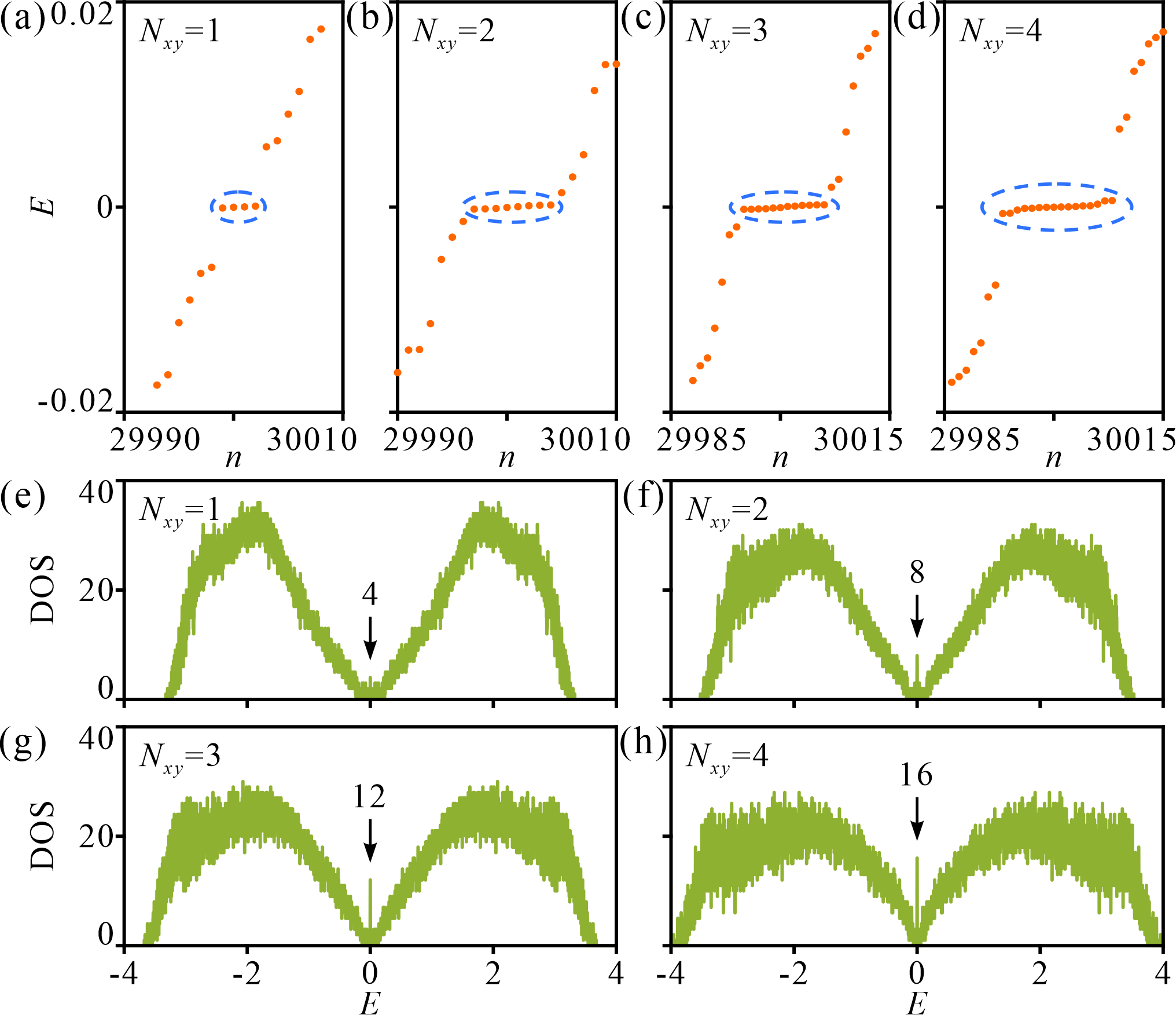}%
\caption{\justifying (a)-(d) Energy spectrum of $H$ and (e)-(h) DOS at a certain open boundary with $W_\mathrm{max} = $ (a, e) 1.72, (b, f) 2.29, (c, g) 2.72 and (d, h) 3.41. The serial number of each state is represented by $n$. The blue dash ellipses circle the point with zero energy state in (a)-(d). The number of zero energy state is shown by the arrows in (e)-(h).\label{}} 
\end{figure}

The above phase transition in HOTAI is not only proved by the topological invariant but also can be revealed by the higher-order bulk-boundary correspondence. Specifically, the quantized ${N_{xy}}$ strictly equals the number of corner states at each corner of the lattice under open boundary conditions. The energy $E$ of the HOTAI described by Eq. (1) at the four peaks in Fig.1 (d) is shown in Fig. 2(a)-2(d). The HOTAI has $4{N_{xy}}$ states with energy near zero (the energy of the corner states is not strictly equal to zero because of the finte size effect). Because of the chiral symmetry, the energy spectrum is symmetric about $E = 0$. Specifically, every eigenstate $\left| {{\psi _n}} \right\rangle $ with energy ${\varepsilon _n}$ has a chiral partner state  $\Pi \left| {\psi_n} \right\rangle=(1 /\sqrt 2){\left[ {\psi _n^{AC},-\psi _n^{BD}} \right]^T}$ that with opposite energy $ - {\varepsilon _n}$. The density of states (DOSs) are shown in Fig.  2(e)-2(h), in which stepwise increases narrow peaks can be observed at $E = 0$ as ${N_{xy}}$ increases from 1 to 4. To prove that these zero energy states are corner-localized states, we define the corner weight as 
\begin{equation}
{w_\mathrm{corner}} = \sum\limits_{{x_c},{y_c}} {{{\sum\limits_{x,y} {\left| { {\psi _n}(x,y)} \right|} }^2}\exp \left[ { - \frac{{{{\Delta x}^2} + \rho{{\Delta y}^2}}}{\xi }} \right]} 
\label{eq:four}
\end{equation}
where $\Delta x=x-x_c$, $\Delta y=y-y_c$ and $({x_c},{y_c})$ represent the positions of the four corners: $(1,1)$,  $(L_x,1)$, $(1,L_y)$ and $(L_x,L_y)$. $\xi  = 0.001({L_x}^2 + {L_y}^2)$ is the decay length and $\rho  = {\left( L_x/L_y \right)^2}$ is used to balance the size difference between $x$ and $y$ directions. A large $w_\mathrm{corner}$ means the field of the state tends to be distributed at the four corners. Fig. 3(a)-3(d) show $w_\mathrm{corner}$ of the zero-energy states, indicating that they are all localized at the corner. The distribution of the zero-energy states shown in Fig. 3(e)-3(h) also displays the corner-localized characteristics.
 
 \begin{figure}
\includegraphics{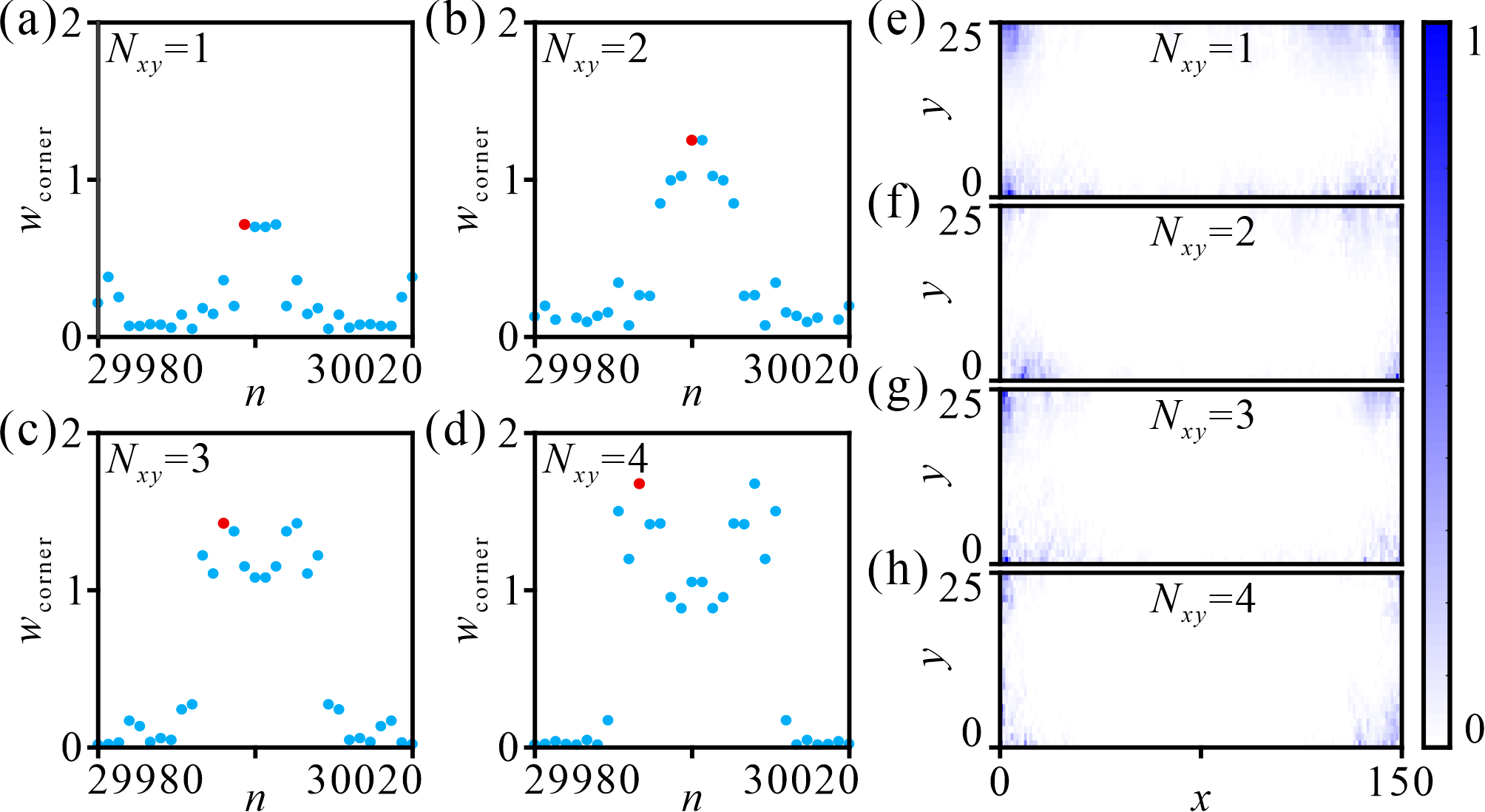}%
\caption{\justifying (a)-(d) The corner weight of eigenstates near the zero-energy for  $W_\mathrm{max} = $ (a) 1.72, (b) 2.29, (c) 2.72 and (d) 3.41. (e)-(h) The field distribution ${\left|  {\psi _n} \right|^2}$ of the state marked by red points in Fig. 3(a)-3(d). \label{}} 
\end{figure}
 
To prove that the four sublattices with different ${v_{xj}}$ indeed respond differently to the disorder. Here we present the phase diagram of the lattice with all ${v_{xj}}$ taking the same value. For example if we set ${v_{xj}} = {\bar v_{x1}} + W$, the successive HOTAI phase transition disappear and the corresponding $H$ represents a ${\mathbb{Z}_2}$-class HOTAI with ${N_{xy}}$ being either 0 or 4 as the phase diagram shown in Fig. 3(a). If ${v_{xj}}$ is changed to ${v_{xj}} = {\bar v_{xj'}} + W$ ($j' = 2,3,4$), the critical value of the topological phase transition point is shifted and a larger disorder is required to drive the phase transition, as shown in Fig. 3(b)-3(d). Therefore, when  ${v_{xj}}$ take different values within a unit cell, disorder will drive discrete regions with individual ${v_{xj}}$ to undergo topological phase transitions gradually as shown in Fig. 4(e). This character is finally reflected in the successive phase transition of the entire model.

\begin{figure}
\includegraphics{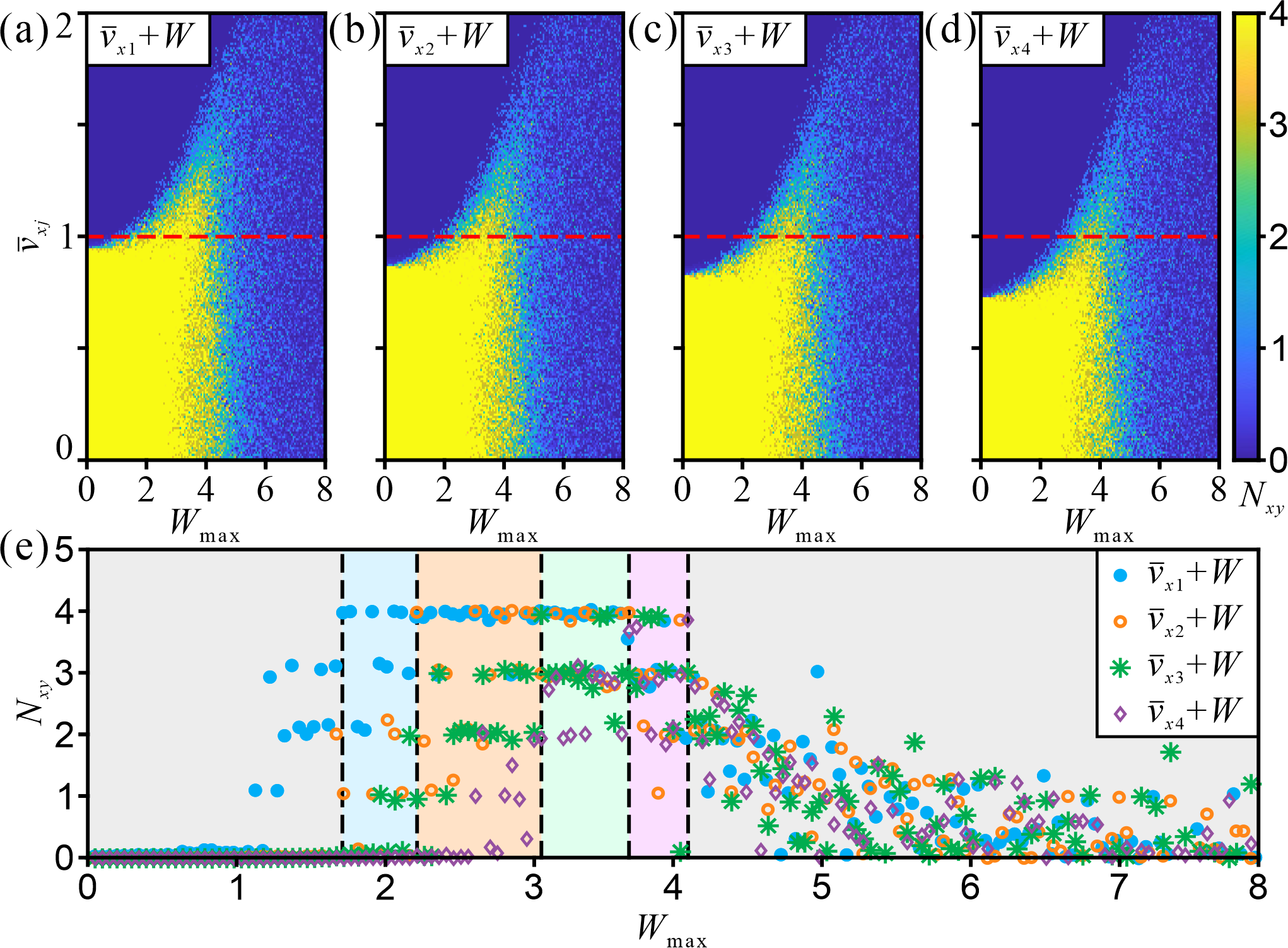}%
\caption{\justifying (a)-(d) Phase diagrams of disordered QTI, in which ${N_{xy}}$ is calculated as a function of ${\bar v_{xj}}$ and $W_\mathrm{max}$. All ${v_{xj}}$ for ${j = 1,2,3,4}$ is chosen to be the same as (a) ${\bar v_{x1}} + W$, (b) ${\bar v_{x2}} + W$, (c) ${\bar v_{x3}} + W$ and (d) ${\bar v_{x4}} + W$, respectively. The red lines capture the phase transition at ${\bar v_{xj}} = 1$. (e) ${N_{xy}}$ versus $W_\mathrm{max}$ along the red line for the above four values taken by ${v_{xj}}$. The light blue (or orange, green, purple) region denote 1 (or 2, 3, 4) ${N_{xy}}$ of the four situations can take the value of 4. \label{}} 
\end{figure}

{\it{Discussion}}.---In conclusion, we here show a successive topological phase transition driven by disorder in HOTI which unveils the connection between disorder strength and diverse non-trivial topological phases. Although we take BBH model as an example, such phase transition phenomenon is expected to widely exist in many other topological insulators such as Su-Schrieffer-Heeger and Harper-Hofstadter-Hatsugai model~\cite{SSH1,SSH2,HHH1}. Besides, our proposed theoretical models can be readily realized in various experimental platforms including acoustics~\cite{AC}, electrical circuits~\cite{LTI2}, photonics~\cite{PH1,PH2} where the long-range couplings and disorders can be introduced straightforwardly. Moreover, it is well known that the influence of disorders on waves behavior significantly different in difference dimensions and in non-Hermitian systems, we also expect further exploration of such successive topological phase transition driven by disorder in higher-dimensional systems such as the 3D quantized octupole insulator~\cite{OI}, in non-Hermitian lattices ~\cite{TAI6,NH1,NH2} and in those with synthetic dimensions~\cite{SD1,SD2,SD3,SD4}.

{\sl Acknowledgments}.
This work was financially supported by The National Key R$\&$D Program of China (2023YFA1407700), Stable Support Program for Higher Education Institutions of Shenzhen (No.20220817185604001), Start-up funding at the Chinese University of Hong Kong, Shenzhen (UDF01002563), National Natural Science Foundation of China (2024SA0007).


\begin{thebibliography}{References}

\bibitem{FW1} M.A. Topinka, B.J. LeRoy, R.M. Westervelt, S.E.J. Shaw, R. Fleischmann, E.J. Heller, K.D. Maranowski, A.C. Gossard, Nature, \textbf{410}, 183-186 (2001).

\bibitem{FW2} A. Patsyk, U. Sivan, M. Segev, M.A. Bandres, Nature, \textbf{583}, 60-65 (2020).

\bibitem{FW3} S.S. Chang, K.H. Wu, S.J. Liu, Z.K. Lin, J.B. Wu, S.J. Ge, L.J. Chen, P. Chen, W. Hu, Y. Xu, H. Chen, D. He, D.Q. Yang, J.H. Jiang, Y.Q. Lu, J.H. Chen, Nat Commun, \textbf{15}, 197 (2024).

\bibitem{AL1} P.W. Anderson, Phys. Rev., \textbf{109}, 1492-1505 (1958).

\bibitem{AL2} F. Evers, A.D. Mirlin, Rev. Mod. Phys., \textbf{80}, 1355-1417 (2008).

\bibitem{AL3} A. Lagendijk, B.v. Tiggelen, D.S. Wiersma, Phys. Today, \textbf{62}, 24-29 (2009).

\bibitem{AL4} E. Abrahams, P.W. Anderson, D.C. Licciardello, T.V. Ramakrishnan, Phys. Rev. Lett., \textbf{42}, 673-676 (1979).

\bibitem{AL5} T. Schwartz, G. Bartal, S. Fishman, M. Segev, Nature, \textbf{446}, 52-55 (2007).

\bibitem{AL6} M. Segev, Y. Silberberg, D.N. Christodoulides, Nat. Photon., \textbf{6}, 197-204 (2013).

\bibitem{AL7} C.A. Condat, T.R. Kirkpatrick, Phys. Rev. Lett., \textbf{58}, 226-229 (1987).

\bibitem{AL8} H. Hu, A. Strybulevych, J.H. Page, S.E. Skipetrov, B.A. van Tiggelen, Nat. Phys., \textbf{4}, 945-948 (2008).

\bibitem{AL9} A. Yamilov, S.E. Skipetrov, T.W. Hughes, M. Minkov, Z. Yu, H. Cao, Nat. Phys., \textbf{19}, 1308-1313 (2023).

\bibitem{AL10} I.S. Burmistrov, I.V. Gornyi, A.D. Mirlin, Phys. Rev. Lett., \textbf{108}, 017002 (2012).

\bibitem{AL11} J. Liu, P.D. Garcia, S. Ek, N. Gregersen, T. Suhr, M. Schubert, J. Mørk, S. Stobbe, P. Lodahl, Nat. Nanotechnol., \textbf{9}, 285-289 (2014).

\bibitem{TI1} M.Z. Hasan, C.L. Kane, Rev. Mod. Phys., \textbf{82}, 3045-3067 (2010).

\bibitem{TI2} J.E. Moore, Nature, \textbf{464}, 194-198 (2010).

\bibitem{TI3} L. Lu, J.D. Joannopoulos, M. Soljačić, Nat. Photon., \textbf{8}, 821-829 (2014).

\bibitem{TI4} B.-C. Xu, B.-Y. Xie, L.-H. Xu, M. Deng, W. Chen, H. Wei, F. Dong, J. Wang, C.-W. Qiu, S. Zhang, Advanced Photonics, \textbf{5}, 036005-036005 (2023).

\bibitem{TI5} D.S. Fisher, Phys. Rev. B, \textbf{50}, 3799-3821 (1994)

\bibitem{TAI1} C.W. Groth, M. Wimmer, A.R. Akhmerov, J. Tworzydło, C.W.J. Beenakker, Phys. Rev. Lett., \textbf{103}, 196805 (2009).

\bibitem{TAI2} J. Li, R.-L. Chu, J.K. Jain, S.-Q. Shen, Phys. Rev. Lett., \textbf{102}, 136806 (2009).

\bibitem{TAI3} Z. Gu, H. Gao, H. Xue, D. Wang, J. Guo, Z. Su, B. Zhang, J. Zhu, Science China Physics, Mechanics \& Astronomy, \textbf{66}, 294311 (2023).

\bibitem{TAI4} E.J. Meier, F.A. An, A. Dauphin, M. Maffei, P. Massignan, T.L. Hughes, B. Gadway, Science, \textbf{362}, 929-933 (2018).

\bibitem{TAI5} H.-C. Hsu, T.-W. Chen, Phys. Rev. B, \textbf{102}, 205425 (2020).

\bibitem{TAI6} I. Mondragon-Shem, T.L. Hughes, J. Song, E. Prodan, Phys. Rev. Lett., \textbf{113}, 046802 (2014).

\bibitem{TAI7} Q. Lin, T. Li, L. Xiao, K. Wang, W. Yi, P. Xue, Nat. Commun., \textbf{13}, 3229 (2022).

\bibitem{TAI8} M. Ren, Y. Yu, B. Wu, X. Qi, Y. Wang, X. Yao, J. Ren, Z. Guo, H. Jiang, H. Chen, X.-J. Liu, Z. Chen, Y. Sun, Phys. Rev. Lett., \textbf{132}, 066602 (2024).

\bibitem{TAI9} S. Stützer, Y. Plotnik, Y. Lumer, P. Titum, N.H. Lindner, M. Segev, M.C. Rechtsman, A. Szameit, Nature, \textbf{560}, 461-465 (2018).

\bibitem{TAI10} G.-G. Liu, Y. Yang, X. Ren, H. Xue, X. Lin, Y.-H. Hu, H.-x. Sun, B. Peng, P. Zhou, Y. Chong, B. Zhang, Phys. Rev. Lett., \textbf{125}, 133603 (2020).

\bibitem{TAI11} X. Cui, R.-Y. Zhang, Z.-Q. Zhang, C.T. Chan, Phys. Rev. Lett., \textbf{129}, 043902 (2022).

\bibitem{QTAI1} C.-A. Li, B. Fu, Z.-A. Hu, J. Li, S.-Q. Shen, Phys. Rev. Lett., \textbf{125}, 166801 (2020).

\bibitem{QTAI2} Y.-B. Yang, K. Li, L.M. Duan, Y. Xu, Phys. Rev. B, \textbf{103}, 085408 (2021).

\bibitem{QTAI3} W. Zhang, D. Zou, Q. Pei, W. He, J. Bao, H. Sun, X. Zhang, Phys. Rev. Lett., \textbf{126}, 146802 (2021).

\bibitem{HOTI1} W.A. Benalcazar, B.A. Bernevig, T.L. Hughes, Science, \textbf{357}, 61-66 (2017).

\bibitem{HOTI2} F. Schindler, A.M. Cook, M.G. Vergniory, Z. Wang, S.S.P. Parkin, B.A. Bernevig, T. Neupert, Sci. Adv., \textbf{4}, eaat0346 (2018).

\bibitem{HOTI3} B. Xie, H.-X. Wang, X. Zhang, P. Zhan, J.-H. Jiang, M. Lu, Y. Chen, Nat. Rev. Phys., \textbf{3}, 520-532 (2021).

\bibitem{RTI1} L. Lin, Y. Ke, C. Lee, Phys. Rev. B, \textbf{103}, 224208 (2021).

\bibitem{LTI1} W.A. Benalcazar, A. Cerjan, Phys. Rev. Lett., \textbf{128}, 127601 (2022).

\bibitem{SSH1}M. Ezawa, Phys. Rev. Lett., \textbf{120}, 026801 (2018).

\bibitem{SSH2} B.-Y. Xie, H.-F. Wang, H.-X. Wang, X.-Y. Zhu, J.-H. Jiang, M.-H. Lu, Y.-F. Chen, Phys. Rev. B, \textbf{98}, 205147 (2018).

\bibitem{HHH1} Y. Kuno, Phys. Rev. B, \textbf{100}, 054108 (2019).

\bibitem{AC} H. Liu, X. Huang, M. Yan, J. Lu, W. Deng, Z. Liu, Phys. Rev. Appl., \textbf{19}, 054028 (2023).

\bibitem{LTI2} Y. Li, J.-H. Zhang, F. Mei, B. Xie, M.-H. Lu, J. Ma, L. Xiao, S. Jia, Phys. Rev. Appl., \textbf{20}, 064042 (2023).

\bibitem{PH1} K. Wang, A. Dutt, K.Y. Yang, C.C. Wojcik, J. Vučković, S. Fan, Science, \textbf{371}, 1240-1245 (2021).

\bibitem{PH2} C. Liu, S. Zhang, S.A. Maier, H. Ren, Phys. Rev. Lett., \textbf{129}, 267401 (2022).

\bibitem{OI} S. Liu, S. Ma, Q. Zhang, L. Zhang, C. Yang, O. You, W. Gao, Y. Xiang, T.J. Cui, S. Zhang, Light Sci. Appl., \textbf{9}, 145 (2020).

\bibitem{NH1} J. Claes, T.L. Hughes, Phys. Rev. B, \textbf{103}, L140201 (2021).

\bibitem{NH2} Q. Lin, T. Li, L. Xiao, K. Wang, W. Yi, P. Xue, Nat. Commun., \textbf{13}, 3229 (2022).

\bibitem{SD1} G. Hu, X. Hong, K. Wang, J. Wu, H.-X. Xu, W. Zhao, W. Liu, S. Zhang, F. Garcia-Vidal, B. Wang, P. Lu, C.-W. Qiu, Nat. Photon., \textbf{13}, 467-472 (2019).

\bibitem{SD2} J. Schulz, J. Noh, W.A. Benalcazar, G. Bahl, G. von Freymann, Nat. Commun., \textbf{13}, 6597 (2022).

\bibitem{SD3} Y. Li, J. Zhang, Y. Wang, H. Du, J. Wu, W. Liu, F. Mei, J. Ma, L. Xiao, S. Jia, Light Sci. Appl., \textbf{11}, 13 (2022).

\bibitem{SD4} F. Guan, X. Guo, K. Zeng, S. Zhang, Z. Nie, S. Ma, Q. Dai, J. Pendry, X. Zhang, S. Zhang, Science, \textbf{381}, 766-771 (2023).

\end{thebibliography}
\end{document}